\documentstyle[snow]{article}
\input epsf
\begin{document}

\onecolumn
\vskip 1.0in

\begin{flushright}
astro-ph/9412046
\end{flushright}

\vskip 1.2in

\begin{center}
{\Huge Quantum Aspects of Gravity}
\end{center}

\vskip 0.8in

{\abstracts{Contributions from the G1 Working Group to appear in
{\it Proceedings of the APS Summer Study on Particle and Nuclear
Astrophysics and Cosmology in the Next Millennium, Snowmass,
Colorado, June 29 - July 14, 1994.}}}

\vskip 0.8in

\begin{center}
{\Large Contents}
\end{center}
\vskip 0.3in

\begin{tabular}{llr}
V. Frolov and L. Thorlacius & QUANTUM ASPECTS OF GRAVITY & 2 \\
{ }&{ }&{ }\\
S.B. Giddings and L. Thorlacius &
INTRODUCTION TO THE INFORMATION PROBLEM & 6 \\
{ }&{ }&{ }\\
A. Strominger & TWO-DIMENSIONAL BLACK HOLES AND THE & { } \\
{ } & INFORMATION PUZZLE & 9 \\
{ }&{ }&{ }\\
V. Frolov & ENTROPY OF BLACK HOLES & 12 \\
{ }&{ }&{ }\\
L. Thorlacius & BLACK HOLE COMPLEMENTARITY & 15 \\
{ }&{ }&{ }\\
D.A. Lowe & CAUSALITY IN STRING THEORY & 18 \\
{ }&{ }&{ }\\
A.O. Barvinski & NONLOCAL EFFECTIVE ACTION AND BLACK HOLES & 22 \\
{ }&{ }&{ }\\
P.C. Argyres & UNIVERSALITY AND SCALING IN BLACK HOLE FORMATION
& 23
\end{tabular}

\setcounter{secnumdepth}{2}

\title{Quantum Aspects of Gravity}

\firstauthors{Valeri P. Frolov}

\firstaddress{CIAR Cosmology Program, Theoretical
Physics Institute, University of Alberta, Edmonton,
Canada T6G 2J1}

\secondauthors{and L\'arus Thorlacius}

\secondaddress{Institute for Theoretical Physics,
University of California, Santa Barbara, CA 93106-4030, USA}

\twocolumn[\maketitle\abstracts{
The reconciliation of gravity theory and quantum physics
is an important goal of theoretical physics and many fundamental
issues can only be settled when a theory of quantum gravity is in
hand.  Superstring theory offers the promise of a unified description
of all interactions including gravity, but many aspects of string
physics remain mysterious.  Theoretical study of the very early
universe and black holes, where Planck scale physics comes into
play, may reveal important clues about fundamental physical law.
}]
\def\epl{E_{\rm pl}}
\def\lpl{l_{\rm pl}}
The G1 Working Group focused on quantum aspects of gravitational
phenomena.  By the nature of the subject, all the talks and
discussions were on theoretical problems and most were on formal
black hole theory, reflecting the current research interests of the
active participants in the Working Group.  In these introductory
words we will discuss some general features of the endeavor which
collectively goes under the heading `quantum gravity' and attempt
to motivate some of the key physical questions that workers in this
area are addressing.  Our remarks will be brief and as there exists
at present no accepted theory of quantum gravity they can only
reflect our personal assessment of the subject.
The general discussion will be followed by
contributions from speakers in the organized lecture sessions of
the Working Group.  These contain accounts of
current research topics but an effort has been made to keep them
accessible to non-experts.

\section{Gravity and Quantum Physics}
Exploring the interface between general relativity and quantum
theory has been one of the great intellectual challenges of
twentieth century physics and will undoubtedly remain one well into
the next millennium.  The measured values of the constants of nature
$c$, $G$, and $\hbar$, which govern the strength of relativistic,
gravitational and quantum effects, indicate that the realm of
quantum gravity is remote indeed.  The characteristic energy and
length scales are the Planck energy,
$\epl=\sqrt{\hbar c^5/G}\sim 10^{19}$ GeV, and the Planck length,
$\lpl=\sqrt{G\hbar /c^3}\sim 10^{-33}$ cm, which are very far out
of reach in present day experiments.

It is nevertheless important to push forward our understanding
at this frontier.  For one thing, Planck scale dynamics may have
subtle effects on physics at lower energies, which cannot be
anticipated without some knowledge of that dynamics.
If our attention is restricted to low-energy physics alone we will
be left with an incomplete theory,
within which many fascinating fundamental questions could
never be resolved.  Furthermore, the very search for underlying
principles can benefit physics in general.  Theoretical efforts
to understand basic physical laws have in the past led to important
insights and new ideas that have had impact on different branches
of physics.

It is difficult to achieve theoretical understanding of fundamental
physical laws without the guiding light of experimental discovery.
The lack of laboratory data has led theorists to rely increasingly
on internal consistency and even mathematical aesthetics as
tools for shaping their theories.  This approach has been
remarkably fruitful, culminating in the development of superstring
theory, which successfully addresses the short distance problems
that undermine a more conventional quantum field theory of gravity.

Superstring models have a number of appealing features.  Gravity is united
with the other known interactions into a single geometric framework,
which can also accommodate multiple generations of chiral fermions
in a natural way.  Unification is widely believed to be a
necessary ingredient in a successful theory of quantum gravity.
On the other hand, our understanding of string dynamics is limited.
The theory is defined in terms of a perturbation expansion around
a classical ground state and there is a multitude of equally valid
ground states to choose from, without any {\it a priori\/}
preference given to one over another.  The formalism is well suited
to the calculation of scattering amplitudes but is not equipped to
address many important issues in quantum gravity, such as initial
conditions, quantum mechanical interpretation, or the cosmological
constant problem.

Despite its shortcomings, string theory remains
the only candidate for a consistent unification of gravity with
quantum physics and a high priority should be placed on obtaining
a deeper understanding of string physics.

While further mathematical developments are clearly called for
it is also important to maintain focus on physical phenomena in
quantum gravity.  In this respect, black holes and the very early
universe  provide a useful testing ground for theoretical ideas.
Conditions extreme enough to bring Planck scale physics into play
are realized in these systems and the various theoretical puzzles of
black hole physics and quantum cosmology provide hurdles for candidate
theories to pass over.  There is also the chance that some
observable features of the Universe can be traced to quantum
gravity effects, either at the earliest epoch or through primordial
black holes.

\section{Black Holes}
In general relativity a black hole is by definition a region of
spacetime which is not in the causal past of future null infinity,
or in less precise words, it is a region
from which signals cannot escape
to the outside world.  Since nothing that occurs inside a classical
black hole can influence events outside, is it even necessary to
develop the physics of black hole interiors?  If the black hole
mass is large compared to the Planck mass then all invariant
features of the geometry outside the event horizon, including the
curvature, are similar to those of flat Minkowski space.
We could hope to get by without invoking quantum gravity at all,
provided we ignore everything that goes on inside the black hole.

There are at least two compelling reasons
why this is not so.  An obvious one is that if we want to use
black holes as tools to learn about strong gravity effects then
we have to go where the action is, {\it i.e.} deep
inside a black hole.  The classical singularity occurs in a region
where the gravitational coupling is strong and quantum effects are
expected to be significant.  A singularity signals a breakdown of
the equations that predict it and it is an interesting question
what replaces the black hole singularity in the quantum theory.

A second reason is that the classical picture of a black hole as
a stable end-result of gravitational collapse is incorrect.  Black
holes emit Hawking radiation and gradually evaporate.  A large
black hole with a smooth external geometry  will eventually reduce
to a small one with strong curvature at the event horizon.  The
final stage of the evaporation can only be described by a quantum
theory and it is unknown at present whether the black hole completely
disappears or whether some strong coupling effects stabilize a
Planck scale remnant.

Black hole evaporation precipitates a serious conflict between
general relativity and quantum physics.  Semi-classical calculations
indicate that a black hole emits black body radiation at the Hawking
temperature $T_h = \hbar c^3/8\pi kGM$, where $M$ is the black hole
mass.  Now imagine matter in a pure quantum mechanical state
undergoing gravitational collapse.  In quantum mechanics a pure
initial state always evolves to a pure final state but if a black
hole is formed then, according to the semi-classical computations,
the final state will contain outgoing thermal radiation and thus be
a mixed state.  The information about the initial pure quantum state
appears to have been lost inside the
black hole.  This paradox has received
a lot of attention in recent years and will be discussed in more
detail in some of the lectures in this volume.  Its resolution will
presumably require us to give up some accepted dogma and this is
often the route to progress.

A related issue concerns thermodynamic properties of black holes.
The laws that govern classical black hole dynamics can be cast in
a form that closely parallels the laws of thermodynamics and the
analogy is strengthened when quantum effects are included.  A black
hole emits blackbody radiation at a temperature which is
proportional to the black hole surface gravity, and appears to
carry an entropy $S=A/4$, where $A$ is the area of the event horizon
in Planck units.  The black hole entropy must be considered
physical if the second law of thermodynamics is to extend to
systems which include black holes, for otherwise one could reduce
the total entropy by sending thermal matter into a black hole.
The ordinary laws of thermodynamics can be understood in terms of
statistical physics and we would like to have a corresponding
explanation of black hole thermodynamics.  An actively pursued
question is how black hole entropy can be given a microphysical
interpretation in quantum gravity.

At the technical level, much of the work on quantum aspects of black
hole physics has been in a semiclassical context, involving
quantized matter in a classical background geometry.  Some progress
has been made in taking into account the back-reaction on the
geometry due to Hawking emission, although so far a controlled
systematic framework is only available in simplified toy models.
It is important to develop this area further but many important
questions cannot be answered within a semiclassical approximation and
will require a more fundamental approach.

We have by no means exhausted the list of interesting questions that
arise in black hole physics.  Let us mention a few more:  1)~In a quantum
theory one expects pair creation of black holes and there is some
controversy as to what the rate of pair creation would be.  2)~Is the
spectrum of black hole states in quantum gravity discrete or
continuous?  The answer to this question is intimately related to the
issues of entropy and information loss mentioned above.  3)~To what
extent do the uniqueness or no-hair theorems on classical black
holes carry over to quantum gravity?  4)~A black
hole carrying electromagnetic charge has a timelike classical
singularity, which becomes visible to observers that enter the
black hole and pass through the so called Cauchy horizon.  The classical
initial value problem breaks down in this case and one may ask how
Cauchy horizons affect quantum mechanical evolution.

\section{Quantum Cosmology}
The other main area where quantum gravity finds application is in
cosmology.  In the hot big bang scenario the Universe expands from a
configuration governed by physics at very high energy, where
quantum effects are expected to be important.  A quantum theory
of cosmology must deal with the earliest epoch in the evolution of
the Universe, where the classical cosmological solution breaks
down at a singularity.  The relevant laws of physics are unknown at
present and the nature of the questions to be answered will depend
on the form these laws take.

If we assume that the Universe evolves from some initial quantum
state we can ask to what extent the theory specifies this initial
state and what observations on the present Universe could in
principle reveal information about it.  Possibly the dynamics of
the subsequent evolution is such as to seek out a particular
configuration for a large class of different initial states, as
for example in inflationary models.  In that case the precise nature
of the initial state is somewhat less important and will at any
rate not be accessible from later observations.

In order to properly address such issues we have to confront some
basic conceptual problems that arise when we attempt to give a quantum
mechanical description of cosmology.  First of all, our Universe
is a unique object whereas the statistical predictions of quantum
mechanics normally apply to ensembles of identically prepared systems.
A closely related concern is the meaning of quantum measurements
when all observers are necessarily part of the quantum system itself.
Another important question is how the classical Universe we inhabit
now, with its irregular matter distribution, evolves from an initial
pure quantum mechanical state.  Recent work on the decoherence of
histories in quantum theory is aimed at addressing some of these
problems.

In many models of quantum cosmology our observable Universe is only
a small piece of a much larger structure, which contains regions of
spacetime that are removed from causal contact with our region, or
even topologically separate `universes'.
We may have to settle for a theory that does not uniquely determine
the initial configuration of our region of spacetime but only assigns
probabilities to different possibilities.  Similarly, the very laws
of low-energy physics in our observable Universe and the number of
observed spacetime dimensions could also be subject to stochastic
rules.  If the probability distribution is strongly peaked around
some given set of low-energy laws and matter configuration we
could still claim that our theory predicts these.  Another
possibility is that the measured values of the constants of nature
and the matter content we observe are only favored by the fact that
our form of intelligent life can only evolve in regions of
spacetime with these properties.  This weak form of the so called
anthropic principle does not give very satisfying answers but is
hard to rule out as a logical possibility.

There is no lack of interesting questions in quantum cosmology.
The challenge for workers in this field is to organize the answers
into a coherent framework and extract physical predictions that
are relevant to the observable world around us.  At present,
quantum cosmology is a speculative enterprise which borders too
closely on the metaphysical for the taste of many physicists.
On the other hand, the inquiry into the origin of the Universe has
always had a strong appeal and physics brings a unique perspective,
rooted in physical observations made on the Universe at large, to
the debate.

\section{The G1 Working Group}
We have touched upon a number of topics that are of current
interest in quantum gravity.  Some of these are discussed further
in the various contributions that follow, and are
based on lectures delivered during the Working Group sessions.
The black hole information paradox has been actively studied in
recent years and a majority of the lectures were concerned with
different aspects of this problem.  The following is a list of
the contributions and the topics covered in each:

S.B.~Giddings and L.~Thorlacius give an introductory review
of the information puzzle.

This is followed by a contribution from A.~Strominger, who first
discusses the information problem in
the simplified context of two-dimensional models and then
describes a possible resolution of the paradox based on
third quantization.

V.~Frolov considers black hole thermodynamics and discusses some
recent efforts to give a dynamical interpretation to the entropy
carried by a black hole.

L.~Thorlacius discusses kinematic requirements for returning
quantum information in Hawking radiation and their possible
implementation in string theory.


Unusual causal properties
of string field theory, which may enable information return,
are described in the contribution by D.A.~Lowe.


The issue of back-reaction on the geometry of spacetime due to
quantum effects is addressed by A.O.~Barvinsky, who discusses an
approach based on a non-local effective action.

The final contribution from the G1 Working Group to these
Proceedings is by P.C.~Argyres and
deals with recent numerical work,
which has exhibited universality and scaling in black hole
formation, reminiscent of scaling in statistical systems.
As it is based on classical Einstein equations this work does not
directly involve quantum aspects of gravity but it hints at
an interesting mathematical structure and may have implications
beyond the classical theory.

\vfill\pagebreak

\setcounter{secnumdepth}{0}
\section{Acknowledgements}
We thank the organizers and the staff of the Summer Study for their
efforts and the participants in the G1 Working Group for lively
discussions.

The work of V.F. is supported in part by NSERC and
that of L.T. is supported in part by NSF grant No. PHY89-04035.

\title{Introduction to the Information Problem}

\firstauthors{Steven B. Giddings}

\firstaddress{Department of Physics, University of
California, Santa Barbara, CA 93106-9530, USA}

\secondauthors{L\'arus Thorlacius}

\secondaddress{Institute for Theoretical Physics,
University of California, Santa Barbara, CA 93106-4030, USA}

\twocolumn[\maketitle\abstracts{
Hawking's black hole information paradox highlights the
incompatibility between our present understanding of gravity and
quantum physics.  The paradox arises in the context of black hole
evolution where infalling matter in a pure initial quantum state
appears to evolve into outgoing thermal radiation.  Its resolution
may involve fundamental information loss, subtle athermal
correlations in the outgoing radiation, or long-lived black hole
remnants.
}]
\setcounter{secnumdepth}{2}
\setcounter{section}{0}
\section{The Puzzle}
\def\mpl{m_{\rm pl}}
\def\tpl{t_{\rm pl}}
\def\Ssl{{\,\raise.15ex\hbox{/}\mkern-10.5mu S}}
The black hole information problem has received considerable attention
as it identifies a serious conflict between quantum mechanics and general
relativity.  To illustrate the problem, consider a gedanken experiment
in which a black hole of mass $M$ is formed in collapse of matter
initially described by a pure quantum state $|\psi\rangle$, with
corresponding density matrix $\rho=|\psi\rangle \langle \psi|$.
An important quantity is the entropy, which for a general density
matrix is given by $S=-$Tr$\rho\ln \rho$, and is zero for this pure
state.  After formation, the black hole decays by emitting
Hawking radiation~[1].  The original semi-classical
calculation involved quantized matter in a classical background black
hole geometry.  It showed the radiation to be approximately thermal and
thus described by a mixed-state density matrix of the form
\begin{equation}
\rho= \sum_n p_n |\psi_n\rangle \langle \psi_n|\ .
\end{equation}
The entropy in the Hawking radiation is nonzero, and can be shown to be
of order $M^2/\mpl^2$ where $\mpl$ is the Planck mass.
This indicates that the final
state is missing information as compared to the initial state, in accord
with the general formula $\Delta I = -\Delta S$ relating information and
entropy.  Hawking argued that the source of the missing information is the
correlation between the particles that come out in Hawking radiation and
those that enter the black hole; without observing the internal state of
the black hole, this information cannot be recovered.

An obvious question is what happens to the information.  There are several
possibilities.  The first is that the black hole completely disappears at
the end of Hawking evaporation.  If the above calculation is to be
trusted, this then means that a pure quantum state has evolved into a
mixed quantum state.  That implies a fundamental loss of information, and
conflicts with the basic principles of quantum mechanics which always
evolves pure states to pure states.  Such a radical conclusion suggests
one consider other options.  One is that some assumptions that go into the
semi-classical calculations are flawed and the information is in fact
gradually released from the black hole over its lifetime and carried in
the Hawking radiation.  Alternately, one might imagine that the
information only escapes after the black hole reaches the Planck mass,
where quantum gravity becomes important and Hawking's calculation fails
in any case.  The resulting outgoing state must, however, contain a huge
amount of information, $I\sim M^2/\mpl^2$, and this information is to be
transmitted using only the remaining energy, $E\sim \mpl$.  The only way
to send a large amount of information with a small amount of energy is to
use a large number of very low energy particles, and this takes a long
time.  A simple estimate~[2] gives a time
\begin{equation}
\tau\sim \left(M\over \mpl\right)^4 \tpl\ ,
\end{equation}
which substantially exceeds the age of the Universe for an initial black
hole formed from a mass as small as that of Pyramid peak.  This means that
in this case one is left with a slowly-decaying black hole remnant.

Each of these three alternatives -- {\it information loss,
information return,} and {\it remnants} -- has been advocated by serious
physicists but in each case serious physical objections have also been
raised.  By considering a sufficiently massive black hole the spacetime
curvature and other local invariant features of the geometry can be made
arbitrarily close to those of flat Minkowski space everywhere outside the
black hole horizon.  It is often claimed that the physics of black hole
evaporation must therefore be governed by low-energy physics and in
particular be independent of Planck scale dynamics.  On the other hand,
each of the three alternatives appears to violate some principle of
low-energy physics and therefore, if one of these scenarios is to be
correct, we expect there must be a loophole in some of the low-energy
arguments, through which Planck scale physics enters in some way.
In what follows we consider each proposal in turn and identify key
theoretical challenges that must be met in order to make it viable.
We will stick to the three basic scenarios outlined above as other
proposed resolutions of the information puzzle usually involve variations
on these themes~[2,3].

\section{Information Loss}
Hawking's proposal that pure states evolve into mixed states due to
gravitational effects manifestly violates the basic quantum mechanical
principle of unitarity.  Furthermore, a general connection between
information and energy indicates that loss of information should imply
violation of
energy conservation.  In particular, it is
natural to expect that virtual processes involving
Planck scale black holes cause information and energy
loss in low-energy physics.
The question is then whether this will be a small effect or one
which severely disturbs the low-energy vacuum.

Hawking attempted to generalize quantum mechanics to allow for
information loss~[4].  He proposed to replace the unitary
$S$ matrix, which maps an initial pure quantum state to a pure final
state, by a superscattering matrix $\Ssl$ which acts on density
matrices rather than state vectors,
\begin{equation}
\rho_{AB}\rightarrow \Ssl_{AB}^{\ CD} \rho_{CD}\ ,
\end{equation}
and can map a pure quantum state to a mixed one.  It was subsequently
argued~[5] that this generalization, at least when
implemented in terms of an evolution equation for density matrices,
violates either locality or energy-momentum conservation.
Hawking's proposal is undermined by virtual black hole creation leading
to unsuppressed Planck scale energy fluctuations.  The challenge to
advocates of the information-loss scenario is, therefore,
to {\it find an explicit description of information loss consistent
with energy conservation}.

\section{Information Return}
Unitarity can be maintained if all information about the quantum
state of the collapsing matter is encoded into the Hawking radiation,
but in this case we have to give up some notions of locality and
causality~[6,7].

Imagine a hapless traveler falling into a large black hole.  From
the traveler's perspective nothing bad happens until near
the singularity long after passing the event horizon.  If the
traveler's information is to be encoded into the Hawking radiation
there must be some physical mechanism that transfers all the
information from the infalling traveler to the outgoing radiation
leaving nothing behind.  Since the Hawking radiation originates
outside the black hole, and the local spacetime curvature is weak
everywhere in its causal past, this mechanism cannot operate in a
local fashion.  In the context of a large black hole the non-local
effects must operate
over long distances in order to reconcile the viewpoints of the
traveler, who feels fine until deep inside the black hole, and of
outside observers, who claim the traveler is completely disrupted
before entering the black hole.  The principle of {\it black hole
complementarity\/}~[7] states that such a reconciliation need not
contradict the known laws of low-energy physics, because in order
to compare measurements made by distant observers to measurements
made inside a black hole one must either be a superobserver outside
spacetime or have knowledge of extreme short-distance physics.

Susskind has suggested that string theory is sufficiently non-local
to enable information return~[8].
Strings are indeed not point-like
objects but na\"\i vely one would only expect non-local effects on
the fundamental string scale.  According to Susskind, the enormous
relative boosts encountered in describing different observers near the
event horizon magnify this short distance non-locality so that it
becomes relevant on macroscopic scales. However, even if the
kinematics of string theory differ sufficiently radically
from point-like theories that information from the contents of a black
hole is in some sense accessible at the horizon, it remains an
open question whether string interactions alter the Hawking
radiation significantly enough
to imprint this information on the outgoing state.
A concrete challenge
in this scenario is thus to {\it find nonlocal dynamics (string or
otherwise) sufficient to transfer the information to the Hawking
radiation.}

\section{Remnants}
If remnants solve the information problem, there must be an infinite
number of remnant species so that they can encompass the information
associated with an arbitrarily large black hole.  These remnants
should have masses $\sim \mpl$.  The obstacle here is that an infinite
variety of remnants with approximately equal masses leads to the
problem of infinite production rates for remnants in ordinary processes.
Examples would be in Hawking radiation from large black holes, in
thermal ensembles, or if the remnants are charged, in Schwinger pair
production in electric fields.  The production rate for a given remnant
species may be incredibly tiny, but the infinite number
of species still gives an infinite total rate.

This logic has been questioned~[9] on the
grounds that it considers remnants
to be essentially similar to elementary particles and in particular to
be described by effective field theory.  A concrete model for remnants
are extremal Reissner-Nordstrom black holes, which are also expected to
have an infinite number of internal states~[10].
Studies of their pair production~[11].
suggest the possible relevance of Planck-scale physics
in computing production rates~[12].
This opens the possibility of finite production.  A challenge to
remnant advocates is then to exhibit the mechanism by which remnants
evade the na\"\i ve estimates, and in particular to {\it provide a
concrete effective description of remnants that avoids the connection
between infinite species and infinite production rates.}

\section{Discussion}
Theorists have already learned a great deal by thinking about
Hawking's paradox.  In particular, our picture of semiclassical black
hole physics, including the back-reaction on the geometry due to
Hawking emission, has been considerably clarified in recent years.
A conclusive resolution of the paradox nevertheless remains elusive.

It is of course quite possible that none of the above proposals
correctly describes the physics of black hole evaporation.
Perhaps some combination of these ideas is closer to the truth
or perhaps some key observation is missing and a completely new
approach required.  The correct scenario may or may not be
determined in the next six years, but understanding the underlying
Planck-scale physics is certain to be an important element of the
physics of the next millennium.

\setcounter{secnumdepth}{0}
\section{Acknowledgements}
We wish to thank the organizers for organizing an interesting and enjoyable
workshop.  The work of S.B.G. is supported in part by DOE grant No.
DOE-91ER40618 and NSF grant No. PHY91-57463 and that of L.T. by NSF grant
No. PHY89-04035.

\vfill\pagebreak

\title{Two-Dimensional Black Holes and the Information Puzzle}

\firstauthors{Andrew Strominger}

\firstaddress{Department of Physics, University of California,
Santa Barbara, CA 93106-9530, USA}

\secondauthors{${ }$}

\secondaddress{${ }$}

\twocolumn[\maketitle
\abstracts{${ }$}
\vspace*{-1.in}
]

\noindent
Several years ago it was realized that two-dimensional methods
provide effective tools for addressing the black hole information
puzzle~[1]. Vigorous application and improvements of these
tools over the last two years have crystallized the puzzle, eliminating
some proposed resolutions and greatly constraining the nature of others.
In the first half of this report I will briefly review some of the work
on these two-dimensional models.  In the second half I will briefly
describe a
possible resolution of the information puzzle that we were led to by
these investigations.
Discussion of other approaches to the puzzle can be found for
example in~[2].

\setcounter{secnumdepth}{2}
\setcounter{section}{0}
\section{Two-dimensional Models}
Two-dimensional dilaton gravity models may be viewed as the
$S$-wave sector of a four-dimensional theory of gravity.  The general
spherically symmetric, four-dimensional line element may be written
\[
ds^2 = g_{ab}(\sigma) d\sigma^a d\sigma^b + e^{-2\phi(\sigma)}
d\Omega^2
\]
where
$
(\sigma^1, \sigma^0) \sim (r,t), ~~g_{ab}
$
is a two-dimensional
metric and $4\pi e^{-2\phi}$ is the area of a two sphere at fixed
$\sigma$ expressed in terms of the dilaton field $\phi$. Thus the
dynamics of four-dimensional $S$-wave gravity are described by a
two-dimensional metric {\it plus} a scalar dilaton.

Models of this general type (when coupled to matter) contain black
holes, Hawking evaporation and, consequently, an information puzzle.
However, they are far simpler than their four-dimensional cousins.
This is in part because two-dimensional quantum
gravity is renormalizable: the short distance problems of quantum
gravity are successfully untangled from the
long-distance information problem.
Even better, it is known that two-dimensional
quantum gravity is equivalent to
conformal field theory and the technology developed in this context over
the last ten years can be fruitfully applied to the black hole problem.

Of course one should always bear in mind the possibility that we are
being led down the garden path.  There is no guarantee that the
resolution of the information puzzle for real four-dimensional black
holes is the same as that for toy two-dimensional black holes.

Two-dimensional models have been improved in steps over the last
four years.  The starting point is the classical
action~[1]
\begin{eqnarray*}
S_{c\ell} = \frac{1}{4\pi} \int d^2\sigma\sqrt{-g}
& &\!\!\!\!\!\!\!\!\!\!\bigl[e^{-2\phi}
(R + 4(\nabla\phi)^2 + 4\lambda^2)\\
& & + \sum\nolimits^N_{i=1} (\nabla f_i)^2\bigr]
\end{eqnarray*}
where $\lambda$ is a dimensionful constant and the $f_i$ are $N$ matter
fields.  Classical solutions of this theory describe formation of black
holes by collapsing matter.

The quantum theory is described to leading order in a $1/N$ expansion by
the effective action~[1]
\[
S_{\rm eff} = S_{c\ell} + \frac{N}{24\pi} \int d^2\sigma
\partial_+\rho\partial_-\rho
\]
in a conformal gauge in which the two-dimensional metric is a
$ds^2 = -e^{2\rho} d\sigma^+ d\sigma^-$.
The last term is deduced from the trace anomaly, and
incorporates the back reaction of the Hawking radiation on the geometry.
Solutions of $S_{\rm eff}$ can be found numerically~[3].  One
finds that black holes shrink due to Hawking emission as expected.
Eventually they reach zero size, a naked singularity appears, and the
computer crashes. While more input is required to evolve beyond this
point, one important lesson can already be gleaned: in large-$N$,
two-dimensional
models the information does not come out of the black hole prior to the
evaporation endpoint.  This was quantified in [4] by direct
calculation of the entropy.

A great improvement of this model was found in references [5,6,7],
where it was noted that equally sensible toy models
for black hole physics can be
obtained by modifying $S_{c\ell}$ with counterterms which vanish at
large radius (large negative $\phi$). These counterterms are constrained
by conformal invariance.  A judicious choice of counterterms leads to a
theory which can be transformed (using field redefinitions) to a free
field theory! Furthermore, in terms of the redefined fields, the
spacetime can be analytically continued both through the origin and
through the singularity.

These improved theories are simple enough that a fully quantum treatment
is feasible.  Notable progress in this direction has been made in
[8], but some issues remain unresolved.  In the remainder of
our discussion we shall concentrate on the large-$N$, semiclassical
behavior.

A semiclassical analysis reveals a disaster lurking in the improved
theories.  Collapsing matter forms a black hole which initially
evaporates as
expected.  Unfortunately, black holes never stop evaporating, even when
the horizon reaches $r=0$! The mass of the spacetime asymptotically
tends to minus infinity. Of course no one believes that
anything like this could happen in the real world.
Evidently these two-dimensional models fail
to provide a faithful model of four-dimensional black hole physics
at (and after) the evaporation endpoint.
Thus while two-dimensional models have been of use for studying
the flow of information in
and out of black holes prior to the endpoint, models developed so far
have not been as useful for studying possible types of endpoint behavior.

Attempts to alleviate this problem have been made by imposing
reflecting boundary conditions at the origin $r=0$ (which corresponds to
a timelike line of constant $\phi$ in the two-dimensional field theory)
[7,9,10].\footnote{The original attempt in [7] had
some technical difficulties which were corrected in
[9,10].} This is a standard problem in conformal field
theory. Interestingly, it turns out that the simplest stable
solution in the present context is highly
non-linear. With these boundary conditions, all incoming matter is
reflected through the origin.  However, there is a threshold for black
hole formation, above which unphysical behavior reappears.  The black
hole never stops evaporating and the mass of the spacetime in the region
exterior to the black hole goes to minus infinity. The reflected pulse
in a sense comes back out, but it does so after the end of time as
measured by an observer exterior to the black hole.  This is a
consequence of distortion of the geometry by the infalling matter.

Indeed both the evaporation endpoint
({\it i.e.} the point at which the black
hole horizon has zero size) and the end of time are (for large infalling
pulses) prior to the future lightcone of the region where the pulse
reaches the origin.  Thus no boundary conditions at the origin can avert
this disaster.

Modifications of the two-dimensional model are needed to
obtain dynamics which are sensible in all regimes.
No fully satisfactory model exists as of this writing. Attempts are being
made to modify the theory as follows [11].

At the evaporation endpoint, the black hole horizon has reached zero (or
Planckian) size.  However, it may have a large interior region.  The
geometry (at least for some spacelike slices) contains a large black
hole interior region (storing lots of information) connected by a
Planckian umbilical cord to the exterior spacetime.  In our discussion
so far, this umbilical cord cannot be broken.  However, if topology
change is allowed in quantum gravity (as we believe to be the case)
eventually the umbilical cord will break and the black hole interior
becomes a baby universe.  This possibility should be incorporated into
the two-dimensional models.

To a string theorist, this process is nothing more than open string
(baby universe) emission from the end of a semi-infinite string (the
original spacetime) and the technology for describing such a process is
at least partially in hand.

\section{Information Retrieval}
Let us assume that black hole evaporation indeed terminates in the decay
of the black hole interior into its own baby universe.  What does this
mean for the issue of information loss?

The answer depends on how one treats the portion of the quantum state
which is carried away by the baby universe. Hawking has proposed that
one simply throw away ({\it i.e.} trace over)
this portion of the state.  This
leads to a theory in which information is irretrievably lost.  Indeed in
this context there is no observable content to the statement that it is
``carried away by a baby universe''.  The information might as well have
been destroyed at a singularity.

A second, inequivalent, proposal [11] is to treat the baby
universes as indistinguishable quantum particles in their own right,
which inhabit a larger ``third-quantized'' Hilbert space.  The
indistinguishability means that one does not just trace separately over
the state of each baby universe, but one symmetrizes by tying together
in all possible ways the traces over all baby universes
created at any time in the entire history of the universe.

This proposal was objected to [12] on the grounds that it
violated cluster decomposition: the symmetrizing over all baby universe
correlates widely separated events.  However, it turns out [13]
that this violation of clustering is physically unobservable, because
the theory decomposes into superselection sectors, in each of which
clustering is valid.  Even better, the scattering is unitary within each
superselection sector! The argument [13] is in essence very
similar to those used earlier in wormhole physics [14], but
with additional subtleties.

The superselection sectors are labeled by the eigenvalues $\alpha_i$ of
the (third-quantized) operators $\phi_i$ which create and destroy a baby
universe in the $i$th quantum state.  The unitary outcome of
gravitational collapse depends on the $\alpha_i$'s, but they cannot be
measured {\it except} by forming black holes and watching them
evaporate.  This requires an enormous number of experiments. Before the
$\alpha_i$'s are known, the outcome of gravitational collapse is
unpredictable, even if the exact solution of quantum
string theory were at hand. Indeed, an averaging over the
unknown $\alpha_i$'s
for the case of a single black hole formation/evaporation
exactly reproduces Hawking's prescription.
Thus, in this proposal,
information is preserved in principle but lost in practice.


\def\NP{{\it Nucl.\ Phys.\ }}
\def\PL{{\it Phys.\ Lett.\ }}
\def\PR{{\it Phys.\ Rev.\ }}
\def\PRL{{\it Phys.\ Rev.\ Lett.\ }}

\title{Entropy of Black Holes}

\firstauthors{Valeri Frolov}

\firstaddress{CIAR  Cosmology Program;  Theoretical Physics
Institute, University of Alberta, Edmonton, Canada T6G 2J1}

\secondauthors{${ }$}

\secondaddress{${ }$}

\twocolumn[\maketitle
\abstracts{${ }$}
\vspace*{-1.in}
]
\setcounter{secnumdepth}{2}
\setcounter{section}{0}
\setcounter{equation}{0}
\setcounter{footnote}{0}
\section{Black-Hole Entropy Problem}
According to the  thermodynamical analogy in  black hole physics,
the
entropy of a black hole  in the Einstein theory of gravity equals
$S^{BH} =A_H
/(4l_{\mbox{\scriptsize{P}}}^2)$,
where   $A_H$   is   the   area   of   a   black   hole   surface
and
$l_{\,\mbox{\scriptsize{P}}}=
(\hbar G/c^3)^{1/2}$   is   the   Planck length
\cite{Beke:72,Beke:73}.

The entropy in  black hole physics plays essentially the same role an
in the usual thermodynamics.  In particular it allows one to estimate
what part of the internal energy of a black hole can be transformed
into work. The generalized second law implies that when a black hole
is a part of a thermodynamical system the total entropy (i.e. the
sum of the entropy of a black hole and the entropy of the surrounding
matter) does not decrease.   The success of the thermodynamical
analogy in black hole physics allows one to hope that this analogy
may be even deeper and it is possible to develop a
statistical-mechanical foundation of black hole thermodynamics.

Thermodynamical and statistical-mechanical definitions of the entropy
are  logically different.
{\it Thermodynamical entropy} $S^{TD}$ is defined by the response of
the free energy $F$ of the system to a change of its temperature:
\begin{equation}\label{1}
dF = -S^{TD} dT .
\end{equation}
(This definition applied to a black hole determines its
Bekenstein-Hawking entropy.)

{\it Statistical-mechanical entropy} $S^{SM}$ is defined as
\begin{equation}\label{2}
S^{SM}=-\mbox{Tr}(\rho \ln \rho ) ,
\end{equation}
where $\rho$ is the density matrix describing the internal state of
the system under consideration.  It is also possible to introduce the
{\it informational entropy} $S^I$ by counting different possibilities
to prepare a system in a  final state with given macroscopical
parameters from different initial states
\begin{equation}\label{3}
S^I =-\sum_{n}{p_n \ln p_n },
\end{equation}
with $p_n$ being the probabilities of different initial states.

In standard case all three definitions give the same answer.

Is the analogy between black holes thermodynamics and the `standard'
thermodynamics complete? Do there exist internal degrees of freedom
of a black hole which are responsible for its entropy?  Is it
possible to apply the statistical-mechanical and informational
definitions of the entropy to black holes and how are they related
with the Bekenstein-Hawking entropy?  These are the questions which
are to be answered.

Historically  first attempts of the statistical-me\-cha\-nical
foundation of the entropy of a black hole were connected with the
informational approach \cite{Beke:73,ZuTh:85}. According to this
approach the black hole entropy is interpreted as ``the logarithm of
the number of quantum mechanically distinct ways that the hole could
have been made"\cite{ZuTh:85}.  The so defined informational entropy
of a black hole is simply related with the amount of information lost
by stretching the horizon, and as was shown by  Thorne and Zurek it
is equal to the Bekenstein-Hawking entropy \cite{ZuTh:85}.

The  dynamical origin of the entropy of a black hole and the relation
between the statistical-mecha\-ni\-cal and Bekenstein-Hawking entropy
have remained unclear. In the present talk I describe some new
results obtained in this direction.

\section{Dynamical Degrees of Freedom}
Calculations  in the framework of the Euclidean approach
initiated by Gibbons and
Hawking\cite{GiHa:76,Hawk:79} relate the entropy of a black hole to
the tree-level contribution of the gravitational action, namely the
action of the Euclidean black hole instanton. In this approach the
entropy of a black hole has pure topological origin, and it remains
unclear whether there exist any real dynamical degrees of freedom
which are responsible for it. The problem of the dynamical origin of
the black hole entropy was intensively discussed recently.  The basic
idea which was proposed is to relate the dynamical degrees of freedom
of a black hole with its quantum excitations. This idea has different
realizations\cite{Hoof:85,FrNo:93,CaTe:93,SuUg:94,GaGiSt:94,BaFrZe:94}.

Here I  discuss the recent proposal  \cite{FrNo:93,BaFrZe:94} to
identify the dynamical degrees of freedom of a black hole with the
states of all fields (including the gravitational one) which are
located inside the black hole.  Such modes (for a non-rotating black
hole) have negative energy.    In the Hawking process, the creation
of a particle outside a black hole (such a particle necessarily has
positive energy) is accompanied by a creation of a corresponding
particle in a mode with negative energy inside a black hole. As a
result these modes  with negative energies are permanently excited
and their state is described by thermal density matrix. Only very
small number of those particles which are created outside a black
hole (external particles) can penetrate the potential barrier and
reach infinity. Namely these particles form the  quantum radiation of
a black hole. All other external particles are reflected by the
potential barrier and fall down into the black hole. During the time
when they are still outside the horizon, the corresponding internal
modes (which are described by a thermal density matrix) give the
contribution to the black hole entropy.

\section{Statistical-Mechanical Entropy}
By averaging over states located outside the black hole one generates
the density matrix of a black hole and can calculate the
corresponding statistical-mechanical entropy $S^{SM}$. The main
contribution to the entropy of a black hole is given by inside modes
of fields located  in the very close vicinity of the horizon.
Contributions of different fields enter $S^{SM}$ additively. The
calculations  give the following result for the contribution of a
chosen field to $S^{SM}$
\begin{eqnarray} \label{4}
	&& S^{SM} = \int {d {\mbox{\boldmath $x$}}}
        \sum_{\lambda}  \mu _\lambda  ({\mbox{\boldmath $x$}})
	s (\beta\omega_\lambda ) , \\
	&&s (\beta\omega ) = {{\beta\omega } \over {e^{\beta\omega
}-1}} -
	\ln (1- e^{-\beta\omega }),
	\end{eqnarray}
where $s (\beta\omega )$ is  the entropy of a single
oscillator with the frequency $\omega$ at temperature $T=1/\beta$.
Here
	\begin{eqnarray}
	\mu _\lambda  ({\mbox{\boldmath $x$}}) = g^{\tau\tau} g
^{1/2}
	[R_\lambda ({\mbox{\boldmath $x$}})]^2
	\end{eqnarray}
is a  phase space density of quantum modes and  $R_\lambda
({\mbox{\boldmath $x$}})$ are spatial harmonics corresponding to the
mode with a collective quantum number $\lambda=(\omega,l,m)$.

The so defined $S^{SM}$ contains a volume divergence, connected with
the integration over the space regions near the horizon and is of the
form\cite{FrNo:93}
\begin{eqnarray}\label{5}
S^{SM}\approx \frac{\alpha}{\pi^2\varepsilon}
\end{eqnarray}
where $\alpha\equiv \frac{\pi^2}{90}\left[
h(0)+\frac{7}{8}h(1/2)+h(1)\right]$, $\varepsilon =(l/r_+)^2$, $h(s)$
is the number of helicities of field of spin $s$, and $l$ is the
proper distance cut-off parameter. One may expect that  quantum
fluctuations of the horizon may provide natural cut-off and make
$S^{SM}$ finite. Simple estimations\cite{FrNo:93} of the cut-off
parameter show that $S^{SM} \approx S^{BH}$.

\section{No-Boundary Wave Function}
Another approach to the problem of dynamical degrees of freedom of a
black hole was proposed in Ref.\cite{BaFrZe:94}. Its basic idea  is
the following. The study  of propagation of perturbations in the
spacetime of a real black hole can be
reduced to the analogous problem for its `eternal version',  (i.e. in
a spacetime of an eternal black hole with the same parameters).  The
space of  physical configurations
of a system including a black hole can be related to the space of
`deformations' of the Einstein-Rosen bridge of the eternal black hole
and
possible configurations of other (besides the gravitational) fields
on it,
which
obey the constrains and preserve asymptotic flatness.   In a
spacetime of an `eternal
version' of a black hole,  perturbations with initial data located on
the inner part of the Einstein-Rosen bridge are propagating to the
future remaining
entirely inside the horizon, and hence the corresponding
perturbations in a `physical'  black hole also always  remain under
the
horizon. That is why these data should be identified
with internal degrees of freedom of a black hole.  A quantum state of
a black hole can be described by a wavefunction   defined as a
functional on the configuration space of deformations of the
Einstein-Rosen bridge. In this representation deformations of the
external and internal parts of the Einstein-Rosen bridge naturally
represent degrees of freedom of matter outside the black hole and
black hole's internal degrees of freedom.  The no-boundary ansatz
(analogous to Hartle-Hawking ansatz in quantum cosmology) singles out
a state which plays the role of ground state of the
system\cite{BaFrZe:94}.  By its construction the no-boundary
wavefunction
 of a black hole is symmetric with respect to the transposition of
the interior and exterior parts of the Einstein-Rosen bridge.  We
call this
property {\em duality}.   For a `real' black hole  formed in the
gravitational
collapse, this exact symmetry is broken. Nevertheless, since there
is a
close relation between physics of a `real' black hole and its
`eternal
version',  the duality of the above type plays an important role and
allows
one, for example, to explain why the approach based on identifying
the
dynamical degrees of freedom of a black hole with its external modes
gives
formally the same answer for the dynamical entropy of a black hole as
our
approach.

For study the fields contribution to the statistical-mechanical
entropy in the one-loop approximation it is sufficient to fix mass
$M$ of a black hole as a parameter in the wave function, and consider
only the  part describing fields perturbations. By tracing over the
external variables one obtains the density matrix of a black hole
and can calculate its statistical-mechanical entropy\cite{BaFrZe:94}.
The result  coincides with (\ref{5}).

\section{Why the Entropy is $A/4$?}
In the above approach (as well as in other dynamical approaches)
different fields give independent contributions to $S^{SM}$. Even if
the cut-off parameter depends on the number of fields in Nature it is
virtually impossible to exclude the dependence of the expression for
$S^{SM}$ on the fields properties. This behavior of $S^{SM}$  differs
this quantity from the universally defined Bekenstein-Hawking
entropy. This was considered by many as the puzzle. In
Ref.\cite{Frol:94}  a simple solution to this puzzle  was proposed.
The Bekenstein-Hawking entropy is analogous to the thermodynamical
entropy, defined by the response of the free energy on the change of
the temperature. The standard prove of the equality of the
statistical-mechanical and thermodynamic entropy requires
commutativity of the operations of tracing over the internal states
of the system and differentiating with respect to the temperature. In
the case of a black hole, where the number and properties of the
internal states depend on the mass of a black hole, which in its turn
depends on the equilibrium temperature, this commutativity property
is not valid. In order to demonstrate this one can write  the
contribution of a chosen field to the free energy of a black hole in
the form, which is similar to (\ref{4})
\begin{eqnarray} \label{6}
	&& F = \int {d {\mbox{\boldmath $x$}}}
        \sum_{\lambda}  \mu _\lambda  ({\mbox{\boldmath $x$}})
	f(\beta\omega_\lambda ) +\ldots, \\
	&&f (\beta\omega ) =
	\ln (1- e^{-\beta\omega )},
	\end{eqnarray}
where ($\ldots$) denotes the terms independent of $\beta$ which do
not contribute to $S^{SM}$. After summation over $l,m$ and
integration over the spatial volume one gets
\begin{equation}\label{7}
F=\int d\omega N(\omega |M,\varepsilon )f(\beta\omega_\lambda )
+\ldots ,
\end{equation}
where $N(\omega |M,\varepsilon )$ is the  density of number of
states.

For the thermal equilibrium the mass $M$ of a black hole is related
with temperature ($\beta \equiv T^{-1}=4\pi M$). That is why for the
calculation of the response $dF$ on the change of the temperature
$dT$ one needs to take into account additional dependence  of $N$ on
$T$. This additional dependence of $N$ on $T$ reflects the fact that
operations $d/dT$ and $\mbox{Tr}$ do not commute for our system. As
the result $S^{TD}$ differs from $S^{SM}$ and one has\footnote{
It is interesting to note that this relation can be used to give a
simple explanation of the entropy `renormalization' procedure, which
was postulated by Thorne and Zurek\cite{ZuTh:85}}  \
$S^{TD}=S^{SM}+\Delta S$.

It is possible to  show that the additional term $\Delta S$ exactly
cancels  the leading (divergent near the horizon) contribution  to
the black hole statistical-mechanical entropy\cite{Frol:94}. As the
result of this cancellation one-loop corrections to the
thermodynamical entropy of a black hole (describing the contribution
to this quantity of the internal dynamical degrees of freedom of a
black hole) are small. This explains the universality (independence
on number and properties of fields) of the expression for the
thermodynamical (Bekenstein-Hawking) entropy of a black hole.

\setcounter{secnumdepth}{0}
\section{Acknowledgements}

This work was supported by the Natural Sciences and Engineering
Research Council of Canada.

\def\NP{{\it Nucl.\ Phys.\ }}
\def\PL{{\it Phys.\ Lett.\ }}
\def\PR{{\it Phys.\ Rev.\ }}
\def\PRL{{\it Phys.\ Rev.\ Lett.\ }}

\title{Black Hole Complementarity}

\firstauthors{L\'arus Thorlacius}

\firstaddress{Institute for Theoretical Physics, University of
California, Santa Barbara, CA 93106-4030, USA}

\secondauthors{${ }$}

\secondaddress{${ }$}

\twocolumn[\maketitle
\abstracts{${ }$}
\vspace*{-1.in}
]
\setcounter{secnumdepth}{2}
\setcounter{section}{0}
\section{Introduction}
In this talk I shall discuss Hawking's information paradox~[1]
solely
from the viewpoint that the information about the initial quantum
state of infalling matter forming a black hole is returned to outside
observers and is encoded in the outgoing Hawking radiation as the black
hole evaporates.  I'll assume that there is no fundamental
information loss and that any stable or long-lived black hole
remnants are finitely degenerate at the Planck scale.  This is
a conservative viewpoint in that it assumes unitarity in all quantum
processes, even when gravitational effects are taken into account,
but it presents a novel view of spacetime physics near an event
horizon.

Imagine a team of technologically advanced observers
studying the formation of a black hole and
its subsequent evaporation from a safe distance.  The observers
prepare a pure quantum state of infalling matter and then
make careful measurements on the Hawking radiation emitted by the black
hole over its entire lifetime.
To determine the final state, our observers will have to
patiently perform an enormous number of such experiments, using an
identically prepared initial state, because only mutually commuting
observables can be measured in any single run.  They also have to be
able to make sophisticated observations of correlations between quanta
emitted at different times in the life of the black hole, for
even if the formation and evaporation process as a whole is
governed by a unitary $S$-matrix the radiation emitted at any given
moment will appear thermal.

Only the region exterior to the black hole event horizon is accessible
in the reference frame of the asymptotic observers.  In this frame the
infalling matter must give up all information about its quantum state to
the outgoing Hawking radiation.  Note that it is not enough for the
information to be imprinted onto the Hawking radiation.  It must also be
removed from the infalling matter as it approaches the event horizon,
for otherwise we would have a duplication of information in the quantum
state in violation of linear quantum mechanics~[2].
A useful analogy is a book set on fire.  All the information initially
contained on the pages can in principle be gleaned from measurements on
the outgoing smoke and radiation but at the end of the day this
information is no longer available in book form.
There is an important difference between the burning book and matter
falling into a black hole.  In the former case it is a well understood
microphysical process which transfers the information from book to
radiation whereas matter in free fall entering a black hole encounters
nothing out of the ordinary upon crossing the event horizon.  The
curvature and other coordinate invariant features of the geometry are
weak there if the black hole is large and the region where strong
gravitational effects are expected is a proper distance of order the
Schwarzschild radius further inside the black hole.

We are thus led to conclude that the physical description of matter
approaching the event horizon differs between the asymptotic and
free fall reference frames by more than is warranted by the usual
behavior of local fields under coordinate transformations.
How serious is this apparent contradiction?
The gravitational redshift between the two frames is enormous; the
relative boost factor grows exponentially with the time measured
in the asymptotic frame and, as 't~Hooft has emphasized~[3],
it becomes much larger than anything that has been achieved
in experiments.  It is therefore legitimate to question whether the
usual Lorentz transformation properties of localized objects correctly
relate observations made in the two frames~[4].

The principle of black hole complementarity~[2] states that there
is no contradiction between having all the information return to
outside observers encoded in the Hawking radiation and having
observers in free fall carry information into a black hole.  The
validity of this principle rests on matter having unusual kinematic
properties at high energy but I will argue that it does not conflict
with low-energy physics.  The basic point is that the apparent
contradiction only comes about when we attempt to compare the physical
description in different reference frames.  The laws of nature are the
same in each frame and low-energy observers
in any single frame cannot establish duplication of information.

\section{The Stretched Horizon}
The evaporation of a large black hole is a slow process
and on a short timescale compared to the black hole lifetime
we can approximate the evolving geometry by a static
Schwarzschild geometry. An outside observer, who is at rest
in Schwarzschild coordinates sees thermal radiation at a temperature
which depends on the spatial position.  Near the black hole this
temperature goes like $T\sim 1/\delta$, where $\delta$ is the proper
distance between the observer and the event horizon.  The high
temperature radiation can be attributed to the acceleration required
to prevent the observer from falling into the black hole, which
diverges in the $\delta\rightarrow 0$ limit.

Our knowledge of the laws governing very high energy physics is
limited and for now I'll only deal phenomenologically with the
region nearest the event horizon where the local temperature is
diverging.  Later in the talk I'll motivate the phenomenological
description by appealing to string theory.
It is well established that, from the point of view of outside
observers, the classical physics of a quasistationary black hole
can be described in terms of a `stretched horizon' which is a
membrane placed near the event horizon and endowed with
certain mechanical, electrical and thermal properties~[5].
The nature of this description is coarse grained in that it is
dissipative and irreversible in time.  One doesn't have to be very
specific about how near the event horizon the stretched horizon is
placed as long as it is close compared to the typical length scale
of the classical problem, which could for example be to describe a
black hole interacting with a companion in a binary.

To go beyond this classical picture we postulate that the coarse
grained thermodynamic description of the classical theory has an
underlying microphysical basis.  The quantum mechanical stretched
horizon is a membrane, carrying microphysical degrees of freedom,
with an area larger than that of the event horizon by one Planck
unit~[2].  The term Planck unit is being used in a loose sense and
simply refers to whatever high-energy scale at which the radical
kinematic behavior, required for returning the information, enters.
In string theory this would be the fundamental string scale which
can be considerably lower than the usual Planck energy.
In order to implement black hole complementarity we
also have to postulate that the membrane has no substance in the
frame of an observer entering the black hole in free fall.

\section{Gedanken Experiments}
It is important to determine whether black hole complementarity
leads to observable duplication of information.
The results of measurements performed inside a black hole are not
available outside the event horizon
so outside low-energy observers cannot establish duplication.
Consider, however, a gedanken experiment~[6] where an
observer first learns the results of measurements made on the
outgoing Hawking radiation, which reveal the quantum state of some
system that was previously sent into the black hole, and then enters
the black hole in order to receive a direct signal from the
same system.

It turns out that it is impossible to carry out this experiment
employing only low energy physics~[6,7].
Outside observers have to carry out correlation measurements
on the outgoing Hawking radiation for a very long time before they
can hope to recover the information about the system that is sent
into the black hole.  An observer who waits outside for that
information and then enters the black hole should receive the
message from the original system before crashing into the
singularity.  The black hole geometry is such that the measurement
on the system must then be made, and the result transmitted, in an
extremely short time after the system passes through the event
horizon.  The timescale is in fact so short that
quanta of frequency $\omega\sim \exp{(M^2)}$, where $M$ is the
black hole mass in Planck units, would have to be
employed to achieve this task~[6]
and the back-reaction on the geometry due to such a high-energy
pulse would be very violent.  Conversely, if the experimenters only
have low energy radiation, {\it i.e.} less than Planck energy, at
their disposal then it will be impossible to transmit the signal
in time for the later observer to receive it before encountering
strong gravitational effects near the singularity.

It is a generic feature of gedanken experiments of this type that
short distance physics enters into their analysis in an essential
way and apparent contradictions with black hole
complementarity can be traced to unwarranted assumptions about
physics beyond the Planck scale.
Another class of experiments involves attempts by external observers
to detect whether quantum information is stored at the stretched
horizon by probing the neighborhood of the event horizon~[6].
Their analysis also requires short-distance physics even for a
large black hole.

\section{The Stretched Horizon in String Theory}
If the ideas presented above are correct then it is essential to
gain understanding of physics at very short distances in order to
resolve the issue of information loss.
String theory is widely believed to provide a consistent
short-distance description of matter and gravity and Susskind
has argued that the kinematic behavior of fundamental strings
is consistent with the requirements of black hole
complementarity~[4].  The basis for this claim is
that zero-point fluctuations of string modes make the
size of a string depend on the time resolution employed~[8].
The shorter the time over which the oscillations of a string are
averaged the larger its spatial extent will appear.

Consider a string configuration in free fall approaching a
black hole event horizon.
An observer at rest far away from
the black hole measures asymptotic time $t$ but because of
the increasing redshift, a unit of asymptotic time corresponds
to an ever shorter time interval,
$\delta\tau\sim\delta t\,\exp{(-t/4M)}$,
in the free-fall frame near the event horizon.
The distant observer is therefore using a shorter and
shorter resolution time to describe the string configuration
and, once it passes within a proper distance of order the
string scale from the event horizon, the string begins to
spread both in the longitudinal and transverse directions~[4].
The longitudinal spread cancels out the
longitudinal Lorentz contraction caused by the black hole geometry.
Meanwhile, the spread in the transverse directions causes the
configuration to cover the entire horizon area in a short time
compared to the black hole lifetime.  In this view,
the stretched horizon is made out of the strings
in the infalling matter which forms the black hole.
On the other hand, the spreading effect is not present
in the free-fall frame, where there is no redshift
to enhance the time resolution, and from the point of view
of an infalling observer there is no stretched horizon, in line
with the principle of black hole complementarity.

Since the stringy stretched horizon is formed from the infalling
matter itself, it efficiently absorbs the quantum information
contained in that matter.  The string spreading process also
thermalizes the stretched horizon~[9].  This comes about
because a fixed time resolution in the asymptotic frame translates
in the free-fall frame into a time-dependent mode cutoff
on the scalar fields, which give the transverse location of the
string.  As time goes on, new modes emerge below the cutoff,
and the random phases of the different modes lead to a classical
stochastic evolution of the scalar fields, which
can be given an interpretation in terms
of a branching diffusion of discrete string bits.
This discussion has entirely been in terms of free string
theory and there are indications that string interactions
significantly enhance the spreading effect~[10].

A crucial remaining question in this approach is how the
thermalized information stored at the stretched horizon gets
imprinted on the outgoing Hawking radiation.  Our present
understanding of interacting string theory is insufficient
to properly address this issue.

To summarize, it appears that black hole complementarity cannot
be ruled out on the basis of known principles
of low-energy physics.  It requires radical kinematic behavior at
high energies, which is not a feature of conventional
\vfill\pagebreak
\noindent
local quantum field theories, but seems
to be realized in string theory.

\setcounter{secnumdepth}{0}
\section{Acknowledgements}
This work was supported in part by NSF Grant No. PHY89-04305.


\title{Causality in String Theory}

\firstauthors{\vspace*{-0.1in} David A. Lowe}

\firstaddress{Department of Physics, University of
California, Santa Barbara, CA 93106-9530, USA}

\secondauthors{${ }$}

\secondaddress{${ }$}

%
%
\def\ss{\sigma}
\def\xm{x^-}
\def\xp{x^+}
\def\xv{\vec x}
\def\yv{\vec y}
\def\ym{y^-}
\def\yp{y^+}
\def\Xv{\vec X}
\def\Yv{\vec Y}
\def\pm{p^-}
\def\pp{p^+}
\def\pv{\vec p}
\def\slc{string light-cone}
\def\lc{light-cone}
\def\dxp{\delta\xp}
\def\dxm{\delta\xm}
\def\dxv{\delta\xv}
\def\al{\alpha}
\def\nv{\vec n}

%
%
\def\PTP{ {\it Prog. Theor. Phys.}}
\def\NP{{\it Nucl. Phys.\ }}
\def\AP{{\it Ann. Phys.\ }}
\def\PL{{\it Phys. Lett.\ }}
\def\PR{{\it Phys. Rev.\ }}
\def\PRL{{\it Phys. Rev. Lett.\ }}
\def\CMP{{\it Comm. Math. Phys.\ }}
\def\JMP{{\it J. Math. Phys.\ }}
\def\JSP{{\it J. Stat. Phys.\ }}
\def\JTP{{\it JETP \ }}
\def\JTPL{{\it JETP Lett.\ }}
\def\JP{{\it J. Phys.\ }}
\def\IJMP{{\it Int. Jour. Mod. Phys.\ }}
\def\Mod{{\it Mod. Phys. Lett.\ }}
\def\NC{{\it Nuovo Cimento \ }}
\def\PRep{{\it Phys. Rep.\ }}

\twocolumn[\maketitle\abstracts{
\vspace*{-0.7in}
The extended structure of fundamental strings gives rise to
a string field theory with unusual causal properties. This
is examined by considering the commutator of string fields
in interacting light-cone string field theory. In general,
this commutator is non-vanishing outside the string analog of the
light-cone. This fact could have important implications for our
understanding of localization of information in quantum
gravity. This talk was based on work done in collaboration with
L.~Susskind and J.~Uglum~[1].
\vspace*{-0.05truein}
}]
\setcounter{secnumdepth}{2}
\setcounter{section}{0}
\setcounter{equation}{0}
\section{Introduction}
Conventional string theory provides us with a perturbative
$S-$matrix.  However, it is of interest to compute more general
observable quantities, such as probability amplitudes at finite
times. Here, we will be primarily interested in the causal
properties of such amplitudes, and we will use string field theory to
try and answer these questions.

In the known covariant formulations of quantized string field theory,
the interactions are nonlocal in the center of mass coordinate
$x^{\mu}$, and, as emphasized in~[2], the initial value
problem breaks down.  Therefore, the theory cannot be canonically
quantized in the conventional way. Fortunately, one is able to fix
light cone gauge, where the interactions are local in the center of
mass coordinate $x^+ = \tau$. This allows a conventional canonical
quantization, and a second--quantized operator
formulation exists~[3-6],
which allows one to perform the kind of
calculations referred to in the preceding paragraph.

Our main concern will be the calculation of the commutator of two
string fields on a flat spacetime background. Further
details may be found in~[1].  In the free theory,
this commutator vanishes outside the ``string light cone''~[7,8],
but it will be shown here  that this is
no longer the case when interactions are included.
The conclusion is that measurements
could detect the information carried by a string state outside the
light cone of the center of mass, but only if these measurements
could be performed with resolution times smaller than the string
scale.  This result supports recent arguments by Susskind
concerning the nature of information in string theory~[9],
and could have important implications for our understanding of
localization of information in a theory of quantum gravity.

\section{Commutator of String Fields}

Let us introduce the light-cone coordinates
\begin{equation}
X^+ = (X^0 + X^{D-1})/ \sqrt{2}, \quad X^-= (X^0-X^{D-1})/ \sqrt{2}
\label{lcoords}
\end{equation}
and parametrize the worldsheet of the string by the variables
$\sigma$ and $\tau$.  Light-cone gauge corresponds to fixing
$X^+(\sigma) = \xp=\tau$.  In the following we will consider open
bosonic strings; the generalization to closed strings and to
superstrings should be similar.  The transverse coordinates are
expanded as
\begin{equation}
\vec X(\sigma) = \xv + 2 \sum_{l=1}^{\infty} \xv_l \cos(l \ss)~.
\label{xtrans}
\end{equation}
In light-cone gauge, the string field is a physical observable and
can be decomposed in terms of an infinite number of component fields.
In the absence of interactions, the string field takes the form~[4,5]
\begin{eqnarray}
&& \Phi(\tau, \xm, \Xv(\sigma)) = \int {{d^{D-2}p} \over
{(2\pi)^{D-1}}} \int_0^{\infty} {{d\pp} \over {2\pp}} \biggl [ \nonumber\\
&&\sum_{ \{\nv_l \} } A(\pp, \pv, \{ \nv_l \} ) e^{i(\pv \cdot \xv -\pp
\xm - \tau \pm)} f_{ \{ \nv_l \} }( \xv_l) \nonumber\\
&&+ h.c. \biggr ] \>.
\label{pdecom}
\end{eqnarray}
Here the light-cone energy of a string state is given by
\begin{equation}
p^- \bigl(\pp, \pv, \{ \nv_l \}\bigr) = {{\pv^{\; 2} + 2\sum_{l, i} l
n_l^i + m_0^2} \over {2\pp}} \>,
\label{lcenergy}
\end{equation}
where $m_0^2$ is the mass squared of the ground state of the string.
For bosonic strings, the ground state is a tachyon and $m_0^2$ is
negative.  In order to effect the light cone quantization and
calculate the commutator, we will regard $m_0^2$ as a positive
adjustable parameter~[7,8].  Of course, this is
inconsistent with Lorentz invariance for the bosonic string, but our
results will be essentially unchanged in the superstring case where
$m_0^2 = 0$.

The $f_{ \{ \nv_l \} }( \xv_l)$ are harmonic oscillator wave
functions given by
\begin{equation}
f_{ \{ \nv_l \} }( \xv_l) = \prod_{l=1}^{\infty} \prod_{i=1}^{D-2}
H_{n_l^i}(x_l^i) e^{-l (x_l^i)^2/ (4\pi) }~,
\label{harmon}
\end{equation}
with $H_{n_l^i}(x_l^i)$ a Hermite polynomial.  The $A$ operators obey
the canonical commutation relations
\begin{eqnarray}
&&[ A(\pp, \pv, \{ \nv_l \} ), A^{\dag}( {\pp}', {\pv}\,', \{ \nv_l \,'
\} )] = 2 \pp \nonumber\\
&&\times (2\pi)^{D-1}
 \delta(\pp-{\pp}') \delta^{D-2}(\pv -
{\pv}\,') \delta_{ \{ \nv_l \}, \{ \nv_l\,' \} }~,
\label{commut}
\end{eqnarray}

A component field is obtained from $\Phi$ by multiplying by the
appropriate wave function (\ref{harmon}) and integrating over the normal
mode coordinates $\xv_l$.  For example, the tachyon field is given by
\begin{eqnarray}
&&T(\tau, \xm, \xv) = \int {{d^{D-2}p} \over {(2\pi)^{D-1}}} \int
{{d\pp} \over {2\pp}} [ \nonumber\\ &&
a_T(\pv, \pp) e^{i[\pv \cdot \xv - \pp \xm  -
\pm \tau]} \nonumber\\ &&
+ a_T^{\dag}(\pv, \pp) e^{-i[\pv \cdot \xv - \pp \xm -
\pm \tau]}] \>,
\label{tacky}
\end{eqnarray}
where $a_T(\pv, \pp) = A(\pv, \pp, \{ {\vec 0} \} )$ and $\pm$ is
given by (\ref{lcenergy}).

Now we want to include a cubic interaction.  The light cone
Hamiltonian becomes $H = H_0 + H_3$, where $H_0$ is the Hamiltonian
for free string field theory and the cubic interaction term $H_3$ is
given by
\begin{eqnarray}
H_3 &=& g \int \Phi_{\al_1}(\vec X_1(\ss)) \Phi_{\al_2}(\vec X_2(\ss))
\Phi_{\al_3}(\vec X_3(\ss)) \nonumber\\ &&
\delta(\sum_{r=1}^3 \al_r) \Delta( \vec
X_1(\ss) - \vec X_2(\ss) - \vec X_3(\ss) ) \nonumber\\
&&\times \mu(\al_1,\al_2,\al_3) \prod_{r=1}^3 d \al_r \prod_{r=1}^3
{\cal D} \vec X_r(\ss) \>,
\label{hthree}
\end{eqnarray}
where $g$ is the open string coupling and $\al_r = 2\pp_r$. The
measure factor $\mu$ and the matrix $\Gamma$ are defined in~[1].

Now that interactions have been included, we wish to determine
whether the commutator of two string fields vanishes when the
arguments of the string fields lie outside the
string light cone~[7,8].  Suppose that
\begin{equation}
[\Phi(\xp_1, \xm_1, \Xv_1(\ss )), \Phi(\xp_2, \xm_2, \Xv_2(\ss ))] =
0
\label{proofone}
\end{equation}
when
\begin{equation}
{1 \over \pi} \int d\ss (X_1(\ss) - X_2(\ss))^2 < 0 \>,
\label{prooftwo}
\end{equation}
where we are using the mostly minus convention for the spacetime
metric.  For fixed $X_2(\ss)$, equation (\ref{proofone}) can be regarded as
a function of $X_1(\ss)$, which vanishes in the entire region in
which equation (\ref{prooftwo}) is satisfied.  Differentiating equation
(\ref{proofone}) with respect to $\xp_1$ and setting $\xp_1 = \xp_2 =
\tau$, one obtains
\begin{equation}
[\dot \Phi(\tau, \xm_1, \Xv_1(\ss )), \Phi(\tau, \xm_2, \Xv_2(\ss ))]
= 0
\label{proofthree}
\end{equation}
in the region in which equation (\ref{prooftwo}) holds.  Here $\dot \Phi$
denotes $\partial\Phi/ \partial \xp$.  At equal light cone times,
equation (\ref{prooftwo}) reduces to
\begin{equation}
{1 \over \pi} \int d\ss (\Xv_1(\ss) - \Xv_2(\ss))^2 > 0 \>.
\label{prooffour}
\end{equation}
Therefore, to prove that the string field commutator does not vanish
identically outside the string light cone, it is sufficient to prove
that equation (\ref{proofthree}) fails to hold when equation
(\ref{prooffour}) is satisfied.

To proceed, note that we can use the Heisenberg equation of motion to
express the field $\dot \Phi$ as
\begin{equation}
\dot \Phi = i[H, \Phi] \>.
\label{Hem}
\end{equation}
Consider now the matrix element
\begin{eqnarray}
\lefteqn{\langle 0| \Phi(\pp_3, \Xv_3(\ss_3)) \,
[ \dot \Phi(\xm_1, \Xv_1(\ss_1)
) , \Phi(\xm_2, \Xv_2(\ss_2)) ] | 0 \rangle} \nonumber\\
&&= i \langle 0| \Phi(\pp_3, \Xv_3(\ss_3)) \, \bigl[ [H, \Phi(\xm_1,
\Xv_1(\ss_1) ) ], \nonumber\\ &&
\Phi(\xm_2, \Xv_2(\ss_2)) \bigr] | 0 \rangle \>,
\label{Matel}
\end{eqnarray}
where all fields are evaluated at $\tau = 0$.  Expanding $H = H_0 +
H_3$, the terms involving $H_0$ all vanish by orthogonality.  This is
a reflection of the fact that the commutator does in fact vanish
outside the string light cone in free string field theory~[7].
The remaining terms can be expressed as
\begin{eqnarray}
\lefteqn{\langle 0| \Phi(\pp_3, \Xv_3(\ss_3)) \,
[ \dot \Phi(\xm_1, \Xv_1(\ss_1)
) , \Phi(\xm_2, \Xv_2(\ss_2)) ] | 0 \rangle } \nonumber\\
&&={{2ig(2\pi)^{5(D-1)/2}} \over
{(2\pi)^3 \pp_3}} \int_0^{\infty} {{d \pp_1}\over {2\pp_1} }
\int_0^{\infty} {{d \pp_2}\over {2 \pp_2} } \nonumber\\
&&\biggl( V(2\pp_1, \Xv_1(\ss_1);
2\pp_2, \Xv_2(\ss_2); -2 \pp_3, \Xv_3(\ss_3)) \nonumber\\
&&-V(-2\pp_1, \Xv_1(\ss_1) ; 2 \pp_2,
\Xv_2(\ss_2) ; -2 \pp_3,
\Xv_3(\ss_3) ) \nonumber\\
&&-V(2\pp_1, \Xv_1(\ss_1) ; -2
\pp_2, \Xv_2(\ss_2) ; -2 \pp_3, \Xv_3(\ss_3) ) \biggr)\nonumber\\
\label{scomm}
\end{eqnarray}
where $V$ is the vertex factor obtained from (\ref{hthree}).  The three
terms represent the three possible kinematical situations, in which
the center of mass of string 3, 2, or 1 lies between the centers of
mass of the other two, respectively.  The $\pp_1$ integral may be
performed by using the $\delta(\sum_{r=1}^3 \al_r)$ factor. Then one
notes that the functional $\delta$-function in (\ref{hthree}) contains a
zero mode piece $\delta^{D-2} ( \sum_{r=1}^3 \al_r \xv_r )$.  Using
one of these delta functions, say for the $x^1$ component, allows the
integral over $\pp_2$ to be performed, and sets
\begin{eqnarray}
\al_2 &=& - \al_3 s~, \nonumber\\
\al_1 &=& (s - 1) \al_3,
\label{longitud}
\end{eqnarray}
where
\begin{equation}
s = {{(x_3^1 - x_1^1)} \over {(x_2^1 - x_1^1)}} \>.
\label{longitudtwo}
\end{equation}

The crucial point to notice is that one is left with a
$(D-3)$-dimensional $\delta$-function requiring the $\xv_r$ to be
collinear, and that each term in (\ref{scomm}) has support on a distinct
ordering of the $\xv_r$ on the line connecting them.  We therefore
find that the commutator of two string fields is in general
non--vanishing outside the string light-cone, $\int d\ss ( \delta
X^{\mu} (\ss) )^2 =0$, when interactions are included. This also
implies the commutator is non--vanishing when the centers of mass of
the strings are spacelike separated.  It should be stressed here that
the non--vanishing of the commutator at spacelike separations has
nothing to do with the fact the bosonic string has a tachyon. The
same will be true in the tachyon--free superstring case.

Of more direct physical interest is the analogous calculation for the
component fields.  For simplicity, we will only consider the
tachyon field, though the generalization to an arbitrary mass
eigenstate is straightforward.  Following the previous line of
reasoning, we compute the matrix element
\begin{equation}
\langle 0| T(\pp_3, \xv_3) \, [ \dot T(\xm_1, \xv_1), T(\xm_2,\xv_2) ]
|0 \rangle
\label{tcomm}
\end{equation}
where, as before, all operators are at time $\tau = 0$.  The vertex
appearing in equation (\ref{tcomm}) is the Mandelstam vertex~[3],
which has the momentum space representation
\begin{eqnarray}
V(\al_r, \pv_r) &=& \delta^{D-2}(\sum_{r=1}^3 \pv_r)
\delta(\sum_{r=1}^3 \al_r) \nonumber\\
&&\times \exp\biggl ( {{\tau_0} \over 2}
\sum_{r=1}^3 { {\pv_r^{\; 2} + m_0^2} \over {\al_r}} \biggr ) \>.
\label{vertex}
\end{eqnarray}
Fourier transforming to coordinate representation, one obtains
\begin{eqnarray}
&&V(\al_r, \xv_r) = \delta^{D-2}(\sum_{r=1}^3 \al_r \xv_r)
\delta(\sum_{r=1}^3 \al_r) \nonumber\\
&&\times\biggl( {{ \al_1 \al_2 \al_3} \over
{8\pi^3  \tau_0}} \biggr)^{(D-2)/2}
 \exp\biggl( {{\tau_0 m_0^2} \over 2} \sum {1\over {\al_r}} \nonumber\\
&&+ { {\al_1 \al_2 \al_3} \over {8 \tau_0}} ({{\xv_1 -\xv_2}\over{\al_3}}
- {{\xv_1-\xv_3}\over {\al_2}})^2 \biggr) \>.
\label{cvertex}
\end{eqnarray}
As was the case for the general string field vertex, equation
(\ref{cvertex}) contains the factor $\delta^{D-2}( \sum_{r=1}^3 \al_r
\xv_r)$, so the result is non--vanishing only when the points $\xv_r$
are collinear. The off-shell vertex corresponds to, say, one tachyon
splitting into two others such that all transverse centers of mass
lie along the same line at equal times. In addition there is a
Gaussian factor depending on the separation of the particles.

Consider a configuration in which $\xv_3$ lies between $\xv_1$ and
$\xv_2$, so that only the first term in equation (\ref{tcomm}) is
non--zero.  A simple calculation then gives
\begin{eqnarray}
\lefteqn{\langle 0| T(\pp_3, \xv_3) \, [ \dot T(\xm_1, \xv_1) , T(\xm_2,\xv_2)
]
|0 \rangle =} \nonumber\\
&& -i e^{i \pp_3 ((1-s) \xm_1 + s \xm_2)} {{\delta^{D-3} (\xv_3 -
\xv_1 - s(\xv_2 - \xv_1))} \over {8 \sqrt{2\pi} (\pp_3)^2 s (1-s)
|x_2^1 - x_1^1|}} \nonumber\\
&&\exp\biggl( - {{s(1-s)} \over {2 \gamma(s)}} (\xv_1
-\xv_2)^2
- m_0^2 {{\gamma(s) (s^2-s+1)}\over {2s(1-s)}} \biggr) \nonumber\\
&&\times \biggl ( {{(2\pi)^2 s (1-s)} \over {\gamma(s)}} \biggr
)^{(D-2)/2} \>,
\label{tcomans}
\end{eqnarray}
where $s$ is given in equation (\ref{longitudtwo}) and
\begin{equation}
\gamma(s) = -[s \log (s) + (1-s) \log (1-s)] \>.
\end{equation}
Note that because of our choice of configuration, $s \in [0, 1]$, and
that $\gamma$ is non--negative.  The matrix element (\ref{tcomans}) depends
on the transverse displacement $|\xv_1 -\xv_2|$ through a Gaussian
factor with variance
\begin{equation}
\sigma^2 = {{\gamma (s)} \over {s(1-s)}} \>.
\label{spread}
\end{equation}
One therefore finds that the matrix element has support over a
distance of order $\sigma^2$ outside the light--cone of the center of
mass.  This spread can be made quite large.  Indeed, for small $s$,
we have
\begin{equation}
\lim_{s \rightarrow 0} {{\gamma (s)} \over {s(1-s)}} \sim -\log (s)
\>,
\end{equation}
so for $s \sim \exp \bigl ( -(\xv_1 - \xv_2)^2 \bigr )$, the matrix
element is appreciable.  This can always be achieved by choosing
$x_3^1$ sufficiently close to $x_1^1$.

The question is whether one is able to resolve this information in
practice.  To get an estimate of how quickly the matrix element is
oscillating in light cone time, we can calculate the matrix element
\begin{equation}
\langle 0| T(\pp_3, \xv_3) \, [\ddot T(\xm_1, \xv_1), T(\xm_2,\xv_2)]|0\rangle
\>,
\label{tcomdd}
\end{equation}
and divide by the matrix element (\ref{tcomm}).  This is proportional to
the frequency of oscillation.  To do this carefully, we must multiply
both (\ref{tcomm}) and (\ref{tcomdd}) by a slowly varying function $f$ and then
integrate over $\xv_2, \ldots, \xv_{D-2}$ to eliminate the
$\delta^{D-3}(\sum \al_r \xv_r)$ factors.  Performing this
calculation leads to the following oscillation time scale
\begin{equation}
\delta t \sim {{\pp_3 s} \over {(\xv_2-\xv_1)^2 \log^{-2} (s) + (D -
2 + m_0^2)}}
\label{tres}
\end{equation}
valid for $s \rightarrow 0$.  For the case of interest, $s = \exp
\bigl ( -(\xv_1 - \xv_2)^2 \bigr )$, this becomes
\begin{equation}
\delta t \sim {{\pp_3 \exp \bigl ( -(\xv_1 - \xv_2)^2 \bigr )} \over
{(\xv_1 - \xv_2)^{-2} + (D - 2 + m_0^2)}} \>.
\label{trestwo}
\end{equation}
The conclusion is that in order to observe the spread of information
over more than a string length, one must perform measurements
involving time scales much smaller than the string time.

The information content of these matrix elements exhibits precisely
the same type of diffusive behavior as was described in~[9],
and which was argued to provide a possible resolution
of the black hole information paradox.  Under conditions relevant to
strings propagating near the horizon of a very massive black hole,
the spread of the matrix elements (\ref{tcomm}) can become arbitrarily
large.  The calculation presented above is relevant because the near
horizon geometry can be approximated by Rindler space, which is
simply a section of flat Minkowski space.  Suppose $T(\tau, \pp_1,
\xv_1)$ and $T(\tau, \pp_3, \xv_3)$ represent two strings, closely
separated in the transverse coordinates, falling together toward the
horizon $\tau = 0$ of a black hole of mass $M$.  As a function of
Schwarzschild time $t$, the radial momentum $P$ of a string, measured
by a static Schwarzschild observer, is given at late times by
\begin{equation}
P (t) = e^{t/4M} \pp \>,
\label{Rmoment}
\end{equation}
where $\pp$ is the (conserved) longitudinal momentum of the string.

We would like to sample string 1 with a string near the horizon which
is very far from string 1 in the transverse coordinates, and which
has radial momentum which is small compared to that of string 1 or 3.
To that end, we will choose $P_2 (t) = P_3(0)$.  This means we must
choose $\pp_2$ as a function of time, given by $\pp_2 (t) = e^{-t/4M}
P_3 (0)$.  For the arguments of these fields, the parameter $s$
appearing in the matrix element (\ref{tcomans}) is given by
\begin{equation}
s = {{\pp_2 (t)} \over {\pp_3}} = e^{-t/4M} \>.
\label{seqn}
\end{equation}
One thus finds that the spread of the Gaussian is given by $\sigma^2
\sim {t \over {4M}}$.  This diffusive behavior is the same as that
found in [9,10] for the mean square transverse radius of
a string.

\vfill\pagebreak

\setcounter{secnumdepth}{0}
\section{Acknowledgements}
This work was supported in part by NSF grant No. PHY91-16964.

\title{Nonlocal Effective Action and Black Holes}

\firstauthors{\vspace*{-0.2in} A. O. Barvinski}

\firstaddress{\vspace*{-0.05in} Nuclear Safety Institute, Bolshaya
Tulskaya 52, Moscow 113191, Russia\break
Lebedev Research Center in Physics,
Leninsky Prospect 53, Moscow 117924, Russia}

\secondauthors{${ }$}

\secondaddress{${ }$}

\twocolumn[\maketitle
\abstracts{${ }$}
\vspace*{-1.2in}
]
\setcounter{secnumdepth}{2}
\noindent
Progress in understanding the quantum physics of black holes and
gravitational collapse to an essential extent depends on the issue of
singularities in the quantum gravitational domain. While classical
Einstein equations typically lead to the development of spacetime
singularities and, thus, give rise to the problem of information
loss, the inclusion of quantum effects can drastically change the
classical picture~[1] due to the back reaction of the vacuum
polarization and particle creation. In certain situations the back
reaction damps the spacetime singularities and even makes the
curvature invariants uniformly bounded
throughout the whole spacetime~[1].
This, in particular, qualitatively changes the setting
of the information problem in quantum gravity.

A direct way to the
quantum back-reaction problem is the formalism of the effective action
and effective equations~[2]. Their calculation cannot be done
explicitly, but one can develop approximation schemes
appropriate to the setting of the quantum gravitational problem. One
such scheme, motivated by the potential boundedness of the
spacetime curvature, is the covariant expansion in powers of
curvatures. A local version of this expansion, known as the
Schwinger-DeWitt technique~[2,3], relies on the smallness
of both the curvatures and their spacetime derivatives and breaks down
in massless theories. In the series of papers~[4-6] this
expansion was generalized to the case of arbitrarily high spacetime
gradients of curvature field strengths. This results in the nonlocal
expansion of the effective action in powers of curvatures with
coefficients -- nonlocal covariant form factors. The systematic
study of these form factors within the semiclassical expansion
theory allows one to reproduce the known conformal (anomalous) part
of the effective action and also obtain its previously unknown
conformally-invariant part unrelated to
local conformal anomalies~[4,5].
This, in particular, allows one to obtain the one-loop vertex form
factors for the theory of the most general type including arbitrary
matter and gravitational fields~[5].

Beyond the
semiclassical loop expansion, the analysis of the covariant
perturbation theory shows that important observable characteristics of
collapsing gravitational systems require only the knowledge of certain
asymptotic limits of the spectral weights of the above form factors.
Numerical values of these limits may have a nonperturbative nature and
can be kept as fundamental parameters of the theory determining the
qualitative behavior of the system. Within such a model-independent
approach to quantum gravity theory~[7,8] it was, in
particular, shown that in the self-consistent problem the asymptotic
flatness of spacetime requires at most logarithmic low-energy
behaviour of the nonlocal form factors, the coefficients of this
logarithmic behaviour determining the energy flux through future
null infinity~[8]. It has been also demonstrated that the
stable component of the Hawking radiation, which plays an
important role in gravitational collapse and the problem of
information loss, begins at cubic order in the effective action
expanded in powers of the curvatures~[8]. This important
contribution, which is currently under study, promises to
maintain the future asymptotic flatness in gravitational systems due
to nontrivial cancellations in nonlocal form factors and also
generates a qualitatively new phenomenon -- coherent gravitational
waves radiated by these systems entirely due to quantum
gravitational effects~[9].


\hyphenation{re-para-metri-za-tion}

\title{Universality and Scaling in Black Hole Formation}

\firstauthors{Philip C. Argyres}

\firstaddress{School of Natural Sciences, Institute for
Advanced Study, Princeton, NJ 08540, USA}

\secondauthors{${ }$}

\secondaddress{${ }$}

\twocolumn[\maketitle\abstracts{
\vspace*{-0.5in}
This is a brief review of numerical work of Refs.\ [1--3]
on the formation of small-mass black holes in classical general
relativity.  Their results are described from a point of view which
treats Einstein's equations as generating a renormalization group
(RG) flow on the space of initial conditions.
}]
\setcounter{secnumdepth}{2}
\setcounter{section}{0}
\setcounter{equation}{0}
\noindent
The recent intriguing numerical results of Choptuik~[1],
Abrahams \& Evans~[2], and Evans \& Coleman~[3],
indicate a connection between classical general relativity and
scaling at critical points in statistical systems.  The purpose
of this talk is to briefly review the results of the above-mentioned
authors, and, along the way, to make a few obvious remarks on the
relation of scaling solutions in general relativity to scaling
in statistical systems.

Choptuik simulated the collapse of spherical wave-packets of
a massless scalar field $\phi$ minimally coupled to gravity.
Abrahams \& Evans simulated the axial collapse of gravity
waves, while Evans \& Coleman examined the spherical collapse
of a relativistic perfect fluid (satisfying $p = {1\over 3} \rho$).
In all three cases it was found that for initial conditions
depending on a generic parameter $s$, there exists a
critical value $s^*$ of this parameter above which no black hole
forms ({\it i.e.}, all the incoming mass
is eventually scattered back to spatial infinity, and the
space-time settles down to flat Minkowski space), while for
$s<s^*$ a black hole forms with a mass $M_{\rm bh}$ satisfying
the universal relation $M_{\rm bh} \sim (s^*-s)^\gamma$, with
$\gamma=0.37\pm0.01$.  This relation holds only for $s$
sufficiently close to $s^*$, and is universal in the sense
that the exponent $\gamma$ does not depend on the shape of the
initial wave packet (the initial conditions) or on the type
of matter or symmetry of the problem (since all three groups
find the same exponent).  Also, Choptuik has found the same
behavior for non-minimally coupled and massive scalar fields
in spherical geometries~[4].

Before discussing this universality more critically, let me
describe in more detail Choptuik's simulation~[1].  Consider
the initial condition consisting of a spherical shell
of scalar field with compact support inside a radius $r_0$, with total
mass $M_0$.  We will let the profile of the scalar field and
its first time derivative be arbitrary but fixed, and will
make a one-parameter family of initial conditions by varying
$r_0$, {\it i.e.}, by translating the scalar wave-packet radially.
A massless scalar field minimally coupled to gravity
is scale-invariant: an overall rescaling of the
coordinates will take one solution into another. So, the
relevant scale-invariant parameters to measure are
$s = r_0/(2M_0)$ and $M_{\rm bh}/M_0$.  Then, qualitatively,
Choptuik's results are sketched in Fig.\ 1, where
	\begin{equation}\label{eqone}
	{M_{\rm bh} \over M_0} \sim
	(s^* -s)^{(0.37)}
	\end{equation}
for $s$ less than but close to $s^*$.  Note that for $s>>s^*$
the initial data is in the weak-coupling regime, since however
wild the shape of the scalar wave-packet, by taking $s$ large
enough and rescaling the radial and time coordinates we see that
the initial data can be made as close to flat space as we like.
In this limit it has been shown analytically that
no black hole forms~[5].  In the opposite extreme,
when $s\leq1$ then all the mass is inside its Schwarzschild
radius and a black hole with mass $M_{\rm bh} = M_0$ is
inevitable.
\begin{figure}[hbtp]
\begin{center}
\leavevmode\epsfxsize=7cm\epsfbox{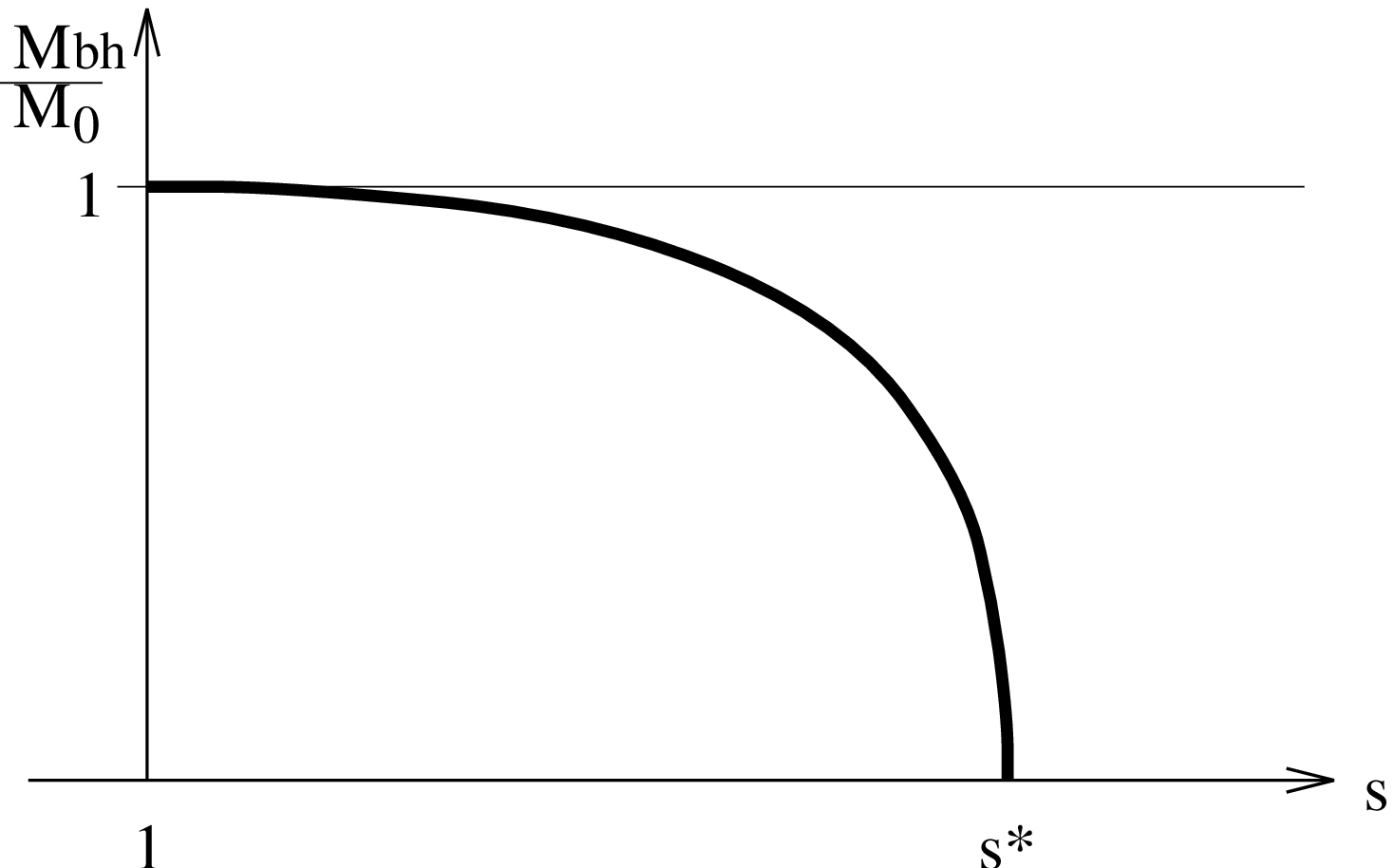}
\end{center}
\caption{}
\end{figure}

The formation of a zero-mass black hole in
the strongly-coupled regime at $s=s^*$ is strongly reminiscent of
the ferromagnetic second-order phase transition where a
spontaneous magnetization arises below a critical temperature.
Taking this analogy seriously suggests that the value of the
critical exponent $\gamma=0.37\pm0.01$, however, must be taken with
a grain of salt, since it has been determined numerically by fitting to
the curve in Fig.\ 1 only over two or three $e$-foldings in
$\ln(s^*-s)$.  Experience with fitting exponents in numerical
simulations in other critical systems indicates that this may only give
a determination accurate at the 20\% level: there are often universal
subleading corrections to the scaling relations which tend to
increase the apparent exponent unless one is much closer to the critical
point $s=s^*$.

As further evidence supporting this analogy to second-order
phase transitions, the numerical simulations also show that
solutions approach universal scaling solutions at criticality.
In order to describe these solutions I first need to describe
the coordinate systems for the resulting space-times. Let us
approach the critical point from large $s$, so that globally
space-time is like Minkowski space. Then, for the spherically
symmetric situations we can use dynamical
Schwarzschild coordinates, for which the line element is
	\begin{equation}\label{eqtwo}
	ds^2 = -\alpha^2(r,t) dt^2 + \beta^2(r,t) dr^2
	+ r^2 d\Omega^2.
	\end{equation}
This fixes the gauge up to an arbitrary reparametrization of
the time coordinate ({\it i.e.} an arbitrary positive function
of $t$).
A natural gauge choice is to fix $\alpha(0,t)=1$, which implies
that $t$ measures central proper time.  For an arbitrary gauge,
the central proper time $T$ is defined by
	\begin{equation}\label{eqthree}
	T(t) = \int^t\alpha(0,\tilde t) d\tilde t .
	\end{equation}
(Note that the origin of central proper time is still not
fixed by this definition.)  The spherical symmetry keeps
the metric functions $\alpha$ and $\beta$ from being dynamical
degrees of freedom.
For a scalar field coupled to gravity with spherical symmetry
(Choptuik's case) only the scalar field $\phi(r,t)$ itself is dynamical.
Einstein's equations can be shown to reduce to a single
nonlinear second order partial integro-differential equation
for $\phi$.  Similarly for a perfect fluid with spherical
symmetry (Evans \& Coleman) the single dynamical
degree of freedom can be taken to be the fluid energy density
$\rho(r,t)$.  In the axially-symmetric pure gravity case (Abrahams
\& Evans) the line element is more complicated, and includes a
function $\eta(r,t,z)$ which is the single dynamical degree of
freedom.

We can classify the self-similar (scaling) solutions to Einstein's
equations in these situations by viewing the evolution equations
as generating a RG flow on the space of initial conditions.
This point of view has proven useful in discussions of
``intermediate asymptotics'' in hydrodynamics and scaling
in nonlinear diffusion equations~[6].  We define
the RG transformation with parameter $\Lambda$ to be simply
	\begin{equation}\label{eqfour}
	{\rm RG}_\Lambda: \phi(r,t)
	\rightarrow \phi_\Lambda(r,t) \equiv
	\Lambda^{\beta_\phi} \phi\left({r\over\Lambda^{\alpha_r}},
	{t^*-t\over\Lambda^{\alpha_t}}\right),
	\end{equation}
where I have used the scalar field for illustration.  A self-similar
(scaling) solution of the equations would then correspond to
a fixed point of the RG transformations.  For example, a continuously
self-similar solution is a solution $\phi^*(r,t)$ satisfying
$\phi^*(r,t) = \phi^*_\Lambda(r,t)$ for all $\Lambda$.  Using the
definition (\ref{eqfour}) this has the solution
	\begin{equation}\label{eqfive}
	\phi^*(r,t) = (t^*-t)^{\beta_\phi/\alpha_t}
	\widetilde{\phi^*}\left({r^{\alpha_t}\over
	(t^*-t)^{\alpha_r}}\right),
	\end{equation}
where $\widetilde{\phi^*}$ is a fixed function of one variable.
More generally, we can consider the situation where the solution
is discretely self-similar.  This solution corresponds to a
{\it limit cycle} of the RG flow instead of a fixed point:
$\phi^*(r,t) = \phi^*_{\Lambda_0^n}(r,t)$ for all integers $n$.
Such a limit cycle can always be written as
	\begin{eqnarray}\label{eqsix}
	\phi^*(r,t) = & \sum_n (t^*-t)^{(\beta_\phi/\alpha_t)
	+(2\pi i n/\ln \Lambda_0)} \nonumber \\
        &\times\
	\widetilde{\phi^*_n}\left({r^{\alpha_t}\over
	(t^*-t)^{\alpha_r}}\right),
	\end{eqnarray}
which is now described by the infinite series of functions
$\widetilde{\phi^*_n}$.

The results of the numerical simulations in this language are:
\begin{center}
\begin{tabular}{rl}
scalar field S-wave [1]:\ \ \ 	& $\alpha_r=\alpha_T=1$\\
\ \ \   & $\beta_\phi=0$\\
\ \ \   & $\Lambda_0 = 3.4$ \\
axial gravity-wave [2]:\ \ \ 	& $\alpha_r=1, \alpha_{t,z}=?$\\
\ \ \   & $\beta_\eta=0$\\
\ \ \   & $\Lambda_0 = 0.6$ \\
perfect fluid S-wave [3]:\ \ \  & $\alpha_r/\alpha_T=1$\\
\ \ \   & $\beta_\rho/\alpha_T=-2$\\
\ \ \   & $\Lambda_0 = 0$
\end{tabular}
\end{center}
\noindent Thus, only the perfect fluid collapse gives a simple
fixed point; the others have limit cycles.  Note that
the appropriate scaling variables in the spherically
symmetric cases are the geometric (areal) radial coordinate
$r$ and the central proper time $T$; and that the exponents all
have their naive scaling values.
Fig.~2 presents a heuristic sketch of the limit cycle RG flows.
\begin{figure}[hbtp]
\begin{center}
\leavevmode\epsfxsize=8.5cm\epsfbox{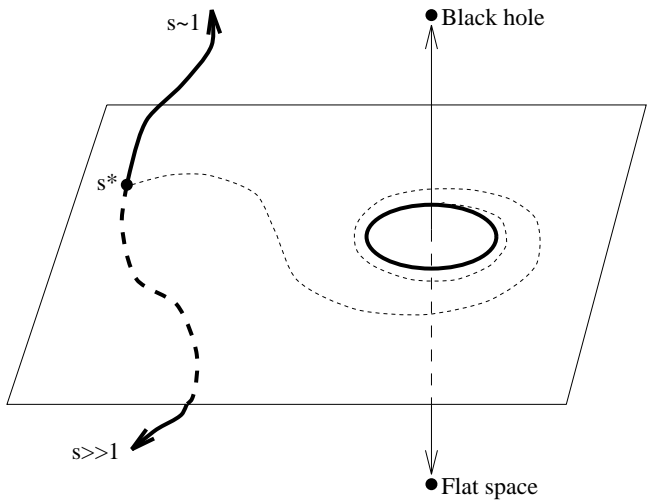}
\end{center}
\caption{A sketch of the (infinite-dimensional) space of
initial conditions.  The plane represents the (co-dimension 1)
critical surface in which the RG flows towards the critical
limit cycle (the circle); off the plane the RG flows either
to the black hole or flat space limits.  The thick line
represents a generic one-parameter family of initial conditions,
while the dotted line in the plane is the RG trajectory of
the $s=s^*$ initial condition.}
\end{figure}

The immediate question presented by these numerical simulations
is how one could analytically compute the critical exponent $\gamma=0.37$.
The most direct approach to this problem is to try to get an
analytic understanding of the critical (scaling) solution, and then
to do linear perturbation theory around this solution to obtain
the exponent(s).  However, since the critical point is in the
strongly-coupled regime, this seems a daunting problem.  In this
regard our experience with critical phenomena in statistical
systems suggests that we look for a simpler (weakly-coupled)
system in the same universality class.
A broader question is whether this critical behavior
persists when we move
away from spherical symmetry or add
more matter fields.  When there is more than one dynamical
degree of freedom, direct computation becomes
\vfill\pagebreak
\noindent
prohibitively difficult, so this question would be best approached
by the type of linear stability analysis mentioned above.

My thanks to M. Bucher, J. Distler, and A. Shapere for helpful
discussions.

\def\PRL{{\it Phys.\ Rev.\ Lett.\ }}
\def\CMP{{\it Comm.\ Math.\ Phys.\ }}

\setcounter{secnumdepth}{0}

\end{document}